\documentclass[a4paper,twocolumn,11pt,unpublished]{quantumarticle}
\pdfoutput=1
\usepackage[utf8]{inputenc}
\usepackage[english]{babel}
\usepackage[T1]{fontenc}

\usepackage{tikz}
\usepackage{lipsum}
\usepackage{graphicx}
\usepackage{lineno,hyperref}
\usepackage{amsmath,amsfonts,amssymb}
\usepackage{amsthm}
\usepackage{array}
\usepackage[caption=false,font=normalsize,labelfont=sf,textfont=sf]{subfig}
\usepackage{textcomp}
\usepackage{url}
\usepackage{graphicx}
\usepackage{epstopdf}
\usepackage{dcolumn}
\usepackage{bm}
\usepackage{braket}
\usepackage{color}
\usepackage{url}
\usepackage{float}
\usepackage{hyperref}
\usepackage{algorithm}
\usepackage{algorithmic}
\usepackage{longtable}
\usepackage{multirow}
\usepackage{makecell}
\usepackage{diagbox}
\usepackage{balance}
\usepackage{soul}
\usepackage{subfig}

\allowdisplaybreaks[2]
\numberwithin{equation}{section}
\modulolinenumbers[5]
\newtheorem{theorem}{Theorem}

\newtheorem{remark}{Remark}

\newtheorem{definition}{Definition}

\makeatletter
\newenvironment{breakablealgorithm}
{
		\begin{center}
			\refstepcounter{algorithm}
			\hrule height.8pt depth0pt \kern2pt
			\renewcommand{\caption}[2][\relax]{
				{\raggedright\textbf{\ALG@name~\thealgorithm} ##2\par}%
				\ifx\relax##1\relax 
				\addcontentsline{loa}{algorithm}{\protect\numberline{\thealgorithm}##2}%
				\else 
				\addcontentsline{loa}{algorithm}{\protect\numberline{\thealgorithm}##1}%
				\fi
				\kern2pt\hrule\kern2pt
			}
		}{
		\kern2pt\hrule\relax
	\end{center}
}
\makeatother

\begin{document}

\title{Distributed Quantum Computing Based on Fixed-point Quantum Search}

\author{Ximing Hua}
\orcid{0000-0001-7163-106x}
\author{Daowen Qiu$^{*,}$}
\orcid{0000-0003-1275-7599}
\affiliation{School of Computer Science and Engineering, Sun Yat-sen University, Guangzhou, 510006, Guangdong, China}
\affiliation{The Guangdong Key Laboratory of Information Security Technology, Sun Yat-sen University, Guangzhou, 510006, Guangdong, China}

\maketitle

\begingroup
\renewcommand{\thefootnote}{*}
\footnotetext{Corresponding author: issqdw@mail.sysu.edu.cn}
\endgroup

\begin{abstract}
Fixed-point quantum search can find target strings without knowing their initial success probability and can be applied to the design of distributed quantum algorithms. This paper makes the following contributions to the integration of distributed quantum computing and the fixed-point quantum search: (1) An inherent relationship between a given Boolean function and its sub-functions is discovered. (2) In particular, a distributed quantum amplitude amplification algorithm based on the fixed-point search is proposed. The algorithm has advantages in the number of qubits and circuit depth, requires no quantum communication, and is simulated by means of Qiskit  to show good noise resistance. (3) Inspired by the proposed algorithm, a general model for applying the fixed-point quantum search to distributed quantum computing is formulated.
\end{abstract}
 

\section{Introduction}

In the  Noisy Intermediate-Scale Quantum (NISQ)  era, the increase of qubits in quantum computers is limited by the development of hardware and the impact of practical environment. These limitations make it difficult to solve large-scale problems by quantum computers. By combining multiple small-scale quantum nodes to jointly solve practical problems, distributed quantum computing has a certain potential to cope with these difficulties \cite{RefKnorzer2026}. To date,  distributed quantum computing has attracted considerable attention from the academic community (e.g., \cite{RefQiu2024,RefBeals2013,RefLeGall2019,RefLi2017,RefQiu2025}). An interesting  method of constructing  oracles in distributed quantum computing has been discussed by Avron et al. \cite{RefAvron2021} in 2021. In recent years,  some basic problems have been solved by using distributed quantum algorithms, including phase estimation \cite{RefQiu2025}, Simon's problem \cite{RefTan2022}, generalized Simon's problem \cite{RefLi2024GS}, integer factorization problem \cite{RefYimsiriwattana2004,RefXiao2023A}, Deutsch-Jozsa problem \cite{RefLiH2025} and generalized Deutsch-Jozsa problem \cite{RefLiH2024DJ}. Notably, a universal method of error correction for distributed quantum computing was discovered in \cite{RefQiu2025}.

Among the numerous algorithms in quantum computing, Grover's algorithm \cite{RefGrover1997} and its generalizations, such as quantum amplitude amplification \cite{RefBrassard2002}, as well as quantum algorithms that employ them as subroutines \cite{RefUtsumi2026,RefGilliam2021}, play an important role. For these algorithms, finding target strings is either the final objective or an essential intermediate procedure. By taking a Boolean function $f:\{0,1\}^n\rightarrow \{0,1\}$ as an example, a string $x$ satisfying $f(x)=1$ is usually referred to as a target string. The initial probability of target strings is generally assumed to be known. However, after $f$ is divided into a number of sub-functions in order to design distributed quantum algorithms, the initial probability of target strings at each node may become unknown.


Qiu et al. \cite{RefQiu2024} proposed distributed Grover's algorithms, and then a precise distributed quantum search algorithm  was given in \cite{RefLiH2024G}. Different from \cite{RefQiu2024,RefLiH2024G}, in this paper, we would employ the fixed-point quantum search \cite{RefGrover2005,RefYoder2014}  to design distributed quantum algorithms, due to its ability to find a target without knowing the  initial probability of target strings in a way. The fixed-point quantum search in \cite{RefGrover2005} leads to the loss of the quadratic speedup, but the fixed-point quantum search algorithm proposed in \cite{RefYoder2014} overcomes this issue,  with its implementation depending on a preset lower bound on the initial probability of target strings.


Based on the above considerations, this paper aims to investigate partial application scenarios of fixed-point quantum search in distributed quantum computing. We propose a distributed quantum amplitude amplification algorithm by using the fixed-point search. In the proposed algorithm, we divide the given Boolean function into multiple sub-functions and derive a useful relationship between a given Boolean function and its sub-functions. Compared with the original algorithm, the proposed algorithm has advantages in the number of qubits and circuit depth. Unlike many other distributed quantum algorithms, it does not require quantum communication. We provide a rigorous proof of the correctness of the algorithm and demonstrate its improved noise tolerance with some simulations conducted by Qiskit. After that, we establish a more general model to characterize the applications of fixed-point quantum search in the distributed quantum computing. Meanwhile, we also illustrate some design strategies for distributed quantum algorithms of three classes problem: the SAT problem, the element distinctness problem, and the graph problems.

The remainder of this paper is structured as follows. In Section  \ref{SEC_Pre}, we revisit the fixed-point quantum search algorithm proposed by Yoder et al \cite{RefYoder2014}. Then, a distributed quantum amplitude amplification algorithm based on the fixed-point quantum search is proposed in Section \ref{SEC_DQAA} with rigorous proofs. Additionally, a simulation of the proposed algorithm and comparisons with related methods are also given in this section. After that, in Section \ref{SEC_APPL_QDC}, we construct a more general model for applying the fixed-point quantum search algorithm in distributed quantum computing and illustrate the applications in many special problems. Finally,  we summarize the main results in the paper and mention a number of problems for further consideration in Section \ref{SEC_Con}.

\section{Revisiting to Fixed-point quantum search algorithm}\label{SEC_Pre}
Fixed-point quantum search aims to solve search problems in which the initial probability of target states is unknown, and in essence, ensures that the probability of searching a target state remains above a prescribed value after a certain number of iterations. The earliest fixed-point quantum search algorithm was proposed by Grover \cite{RefGrover2005} in 2005, but compared with Grover's algorithm \cite{RefGrover1997}, it lost the quadratic speedup. Later, Yoder et al. \cite{RefYoder2014} improved this algorithm in 2014 and proposed another fixed-point quantum search algorithm that preserves the quadratic speedup without knowing the exact initial probability of target states. In this section, we begin with the reviewing of original amplitude amplification algorithm \cite{RefBrassard2002} and finally focus on introducing the fixed-point quantum search algorithm proposed by Yoder et al \cite{RefYoder2014}.

\subsection{Quantum Amplitude Amplification}\label{SEC_QAA}
Quantum amplitude amplification is used to enhance the success probability for a quantum algorithm without involving  measurement. We give an overview of quantum amplitude amplification, and further details can be found in \cite{RefBrassard2002}.

Given a Boolean function $f:\{0,1\}^n\rightarrow \{0,1\}$, let $\mathcal{A}$ be any given quantum algorithm without using any measurement (actually it can be described by a unitary operator), and
$\mathcal{A}$ is applied to the initial state $|0\rangle^{\otimes n}$, denoted by $|\psi\rangle=\mathcal{A}|0\rangle^{\otimes n}$.   The objective is to find out an $x$ with $f(x)=1$. 
Suppose
\begin{equation}
	|\psi\rangle=|\Psi_1\rangle+|\Psi_0\rangle
\end{equation}
where
\begin{equation*}
	|\Psi_1\rangle=\sum_{x:f(x)=1}^{}\alpha_x|x\rangle,
\end{equation*}
and
\begin{equation*}
	|\Psi_0\rangle=\sum_{x:f(x)=0}^{}\alpha_x|x\rangle.
\end{equation*}
Then,  let $a=\langle\Psi_1|\Psi_1\rangle$ denote the initial success probability of searching for an $x$ with $f(x)=1$.

\begin{figure}
	\centering
	\includegraphics[scale=0.4]{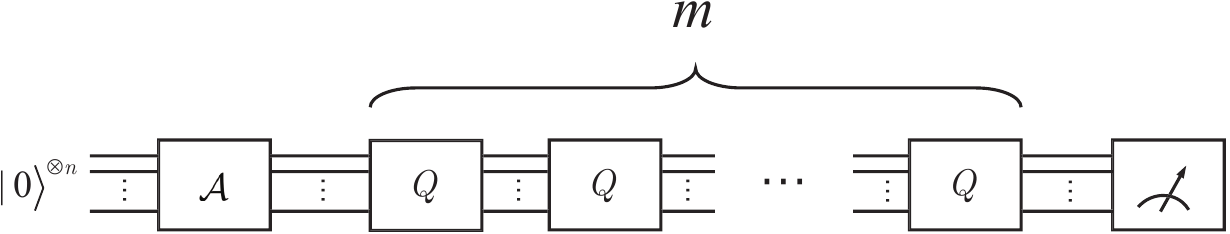}
	\caption{Quantum amplitude amplification.}
	\label{QAA}
\end{figure}

Quantum amplitude amplification is realized by repeatedly applying the iterative operator $Q$ as Fig.\ref{QAA}, and it is defined as follows \cite{RefBrassard2002}:
\begin{equation}\label{EQ_Q}
	Q=-\mathcal{A}S_0\mathcal{A}^{\dagger}S_{f}
\end{equation}
where 
\begin{equation}\label{EQ_Sf}
	S_f|x\rangle=
	\left\{ 
	\begin{array}{lrc}
		-|x\rangle,&f(x)=1,\\
		|x\rangle,&f(x)=0,
	\end{array}	
	\right.
\end{equation}
and
\begin{equation}\label{EQ_S0}
	S_0|x\rangle=
	\left\{ 
	\begin{array}{lrc}
		-|x\rangle,&x=0^n,\\
		|x\rangle,&x\neq 0^n.
	\end{array}	
	\right.
\end{equation}
When $\mathcal{A}=H^{\otimes n}$, the iterative operator is Grover's operator \cite{RefGrover1997}. 

Similar to Grover's algorithm, the number of iterations $m$ is related to $a$.  If $a$ is  known, then  the quantum amplitude amplification is given in the following algorithm \cite{RefBrassard2002}.

\begin{breakablealgorithm}
	\caption{Quantum amplitude amplification}
	\label{QAA_a1}
	\begin{algorithmic}[1]
		
		\REQUIRE {Algorithm $\mathcal{A}$, function $f$, $a$.}
		\ENSURE {$x\in \{0,1\}^n$, with $f(x)=1$.}
		\STATE $|\psi_0\rangle=\mathcal{A}|0\rangle^{\otimes n}$.
		\STATE Apply iterative operator $Q$ to $|\psi_0\rangle$ with $m$ times.
		\STATE Measure the quantum register and obtain  $x$ satisfying  $f(x)=1$.
		\RETURN $x$
		
	\end{algorithmic}
\end{breakablealgorithm}

Algorithm \ref{QAA_a1} and the subsequent algorithms use projective measurements with the associated projectors $\{\Pi_x\}$, where $\Pi_x=|x\rangle\langle x|$ and $x\in\{0,1\}^n$.
Denote $|t\rangle=\frac{1}{\sqrt{a}}|\Psi_1\rangle$.  Then the probability $P_m$ that the outcome $x$ satisfies $f(x)=1$ in Algorithm \ref{QAA_a1}  is actually equal to 
\begin{equation}\label{EQ_Pro_Pm}
	|\langle t|S_m|\psi_0\rangle|^2
\end{equation}
where $S_m=Q^m$, $m$ is the number of iterations. 

To illustrate the above algorithm, we recall the following theorem in  \cite{RefBrassard2002}.

\begin{theorem}\label{TH_QAA}
	Let $\mathcal{A}$ be any quantum algorithm that uses no measurements, and let a Boolean function $f:\{0,1\}^n\rightarrow \{0,1\}$. Suppose $a>0$, and set $m=\lfloor\frac{\pi}{4\arcsin(\sqrt{a})}\rfloor$, where  $0<\arcsin(\sqrt{a})<\pi/2$. If we measure the quantum register in step 3 of Algorithm \ref{QAA_a1}, then the probability of obtaining an $x$ satisfying $f(x)=1$ is at least $\max (1-a,a)$.
\end{theorem}

However, if $a$ is  unknown, then the above algorithm does not work, but in this case,  this problem can be handled instead by utilizing the fixed-point quantum search \cite{RefGrover2005,RefYoder2014},  and we focus on introducing it in the following subsection.

\subsection{Fixed-point Quantum Search Algorithm}\label{Sec_FPQAA}
As noted above, the fixed-point quantum search algorithm was proposed by Grover \cite{RefGrover2005}  and further improved by Yoder et al. \cite{RefYoder2014}. Yoder et al. exploited the properties of Chebyshev polynomials to ensure that, once the number of iterations exceeds a certain threshold, the probability of finding a target state always remains above a prescribed fixed point. In this paper, we would like to employ the algorithm by Yoder et al. \cite{RefYoder2014}, so we introduce  it carefully. 

First we give the following  operators presented in \cite{RefLong1999} and \cite{RefHoyer2000} to implement arbitrary phase rotations.
\begin{equation}\label{EQ_Q_Generlize}
	Q(\mathcal{A},f,\phi,\varphi)=-\mathcal{A}S_0(\phi)\mathcal{A}^{\dagger}S_{f}(\varphi),
\end{equation}
where
\begin{equation}\label{EQ_Sf_Generlize}
	S_f(\varphi)|x\rangle=
	\left\{ 
	\begin{array}{lrc}
		e^{\mathbf{i}\varphi}|x\rangle,&f(x)=1,\\
		|x\rangle,&f(x)=0,
	\end{array}	
	\right.
\end{equation}
and
\begin{equation}\label{EQ_S0_Generlize}
	S_0(\phi)|x\rangle=
	\left\{ 
	\begin{array}{lrc}
		e^{-\mathbf{i}\phi}|x\rangle,&x=0^n,\\
		|x\rangle,&x\neq 0^n.
	\end{array}	
	\right.
\end{equation}
Here, $\mathbf{i}$ denotes the principal square root of $-1$. Let $S_d(\phi)=-\mathcal{A}S_0(\phi)\mathcal{A}^{\dagger}$.   Then 
\begin{equation}\label{EQ_Q_Sd_Sf}
	Q(\mathcal{A},f,\phi,\varphi)=S_d(\phi)S_{f}(\varphi).
\end{equation}
\begin{remark}
	When $\phi=\varphi=\pi$, $Q(\mathcal{A},f,\phi,\varphi)$ in Eq.\eqref{EQ_Q_Generlize} is the iterative operator in \cite{RefBrassard2002} as Eq.\eqref{EQ_Q}.
\end{remark}

The iterative process of fixed-point quantum amplitude amplification \cite{RefYoder2014} can be represented as
\begin{equation}
	S_l=Q(\mathcal{A},f,\phi_l,\varphi_l)...Q(\mathcal{A},f,\phi_2,\varphi_2)Q(\mathcal{A},f,\phi_1,\varphi_1).
\end{equation}
That is to say, let $|\psi_0\rangle=\mathcal{A}|0\rangle^{\otimes n}$ be the initial state and let $|\psi_l\rangle$ be the final state. Then $|\psi_l\rangle=S_l|\psi_0\rangle$. For $r=1,2,...,l$, the angle $\phi_r$ and $\varphi_r$ are defined as follows \cite{RefYoder2014}.
\begin{equation}\label{phi_varphi_r}
	\phi_r=-\varphi_{l-r+1}=2\cot^{-1}\left(\tan(\frac{2\pi r}{2l+1})\sqrt{1-\gamma^2}\right).
\end{equation}
Here, $\gamma^{-1}=T_{\frac{1}{L}}(\frac{1}{\varepsilon})$, $L=2l+1$. For any real number $m$, the function $T_m(x)$ over $[-1,+\infty)$ utilized in this section is defined as follows.

\begin{widetext}
	\begin{equation}\label{EQ_def_T}
		T_m(x)=\left\{\begin{array}{cc}
			\cos(m\arccos(x)),&-1\leqslant x< 1,\\
			\cos(m\arccos(x))=\cosh(m\hspace{0.5mm}\mathrm{arccosh}(x)),& x= 1,\\
			\cosh(m\hspace{0.5mm}\mathrm{arccosh}(x)),& x>1,
		\end{array}\right.
	\end{equation}
\end{widetext}
where $\cosh x=\frac{e^x+e^{-x}}{2}$ and $\mathrm{arccosh}(x)=\ln(x+\sqrt{x^2-1})$.
\begin{remark}
	$\sinh x=\frac{e^x-e^{-x}}{2}$ and $\cosh x=\frac{e^x+e^{-x}}{2}$ are hyperbolic functions, which have similar properties with trigonometric functions. In addition, $\tanh x=\frac{\sinh x}{\cosh x}$, $\coth x=\frac{\cosh x}{\sinh x}$, $\mathrm{sech} x=\frac{1}{\cosh x}$ and $\mathrm{csch} x=\frac{1}{\sinh x}$ are  defined in terms of hyperbolic functions. 
\end{remark}
\begin{remark}
	If $m$ is a positive integer, then $T_m(x)$ is the first kind generalized Chebyshev polynomial \cite{RefRivlin2020} of degree $m$ such that $T_0(x)=1$, $T_1(x)=x$ and for $m\geqslant2$
	\begin{equation}
		T_m(x)=2xT_{m-1}(x)-T_{m-2}(x).
	\end{equation}
\end{remark}

The procedure of the algorithm by Yoder et al.\cite{RefYoder2014} is given as follows.
\begin{breakablealgorithm}
	\caption{Fixed-point quantum amplitude amplification algorithm}
	\label{Al_FPQAA}
	\begin{algorithmic}[1]
		
		\REQUIRE Algorithm $\mathcal{A}$, function $f$, parameter $\delta$, error rate $\varepsilon^2$.
		\ENSURE $x\in \{0,1\}^n$ such that $f(x)=1$.
		\STATE Set $l=\left\lceil\frac{\ln(2/\varepsilon)}{2\delta}\right\rceil$.
		\STATE $|\psi_0\rangle=\mathcal{A}|0\rangle^{\otimes n}$.
		\STATE $|\psi_l\rangle=S_l|\psi_0\rangle$, where $S_l=Q(\mathcal{A},f,\phi_l,\varphi_l)...Q(\mathcal{A},f,\phi_2,\varphi_2)Q(\mathcal{A},f,\phi_1,\varphi_1)$.
		\STATE Measure the register and obtain an  $x$ with $f(x)=1$.
		\RETURN $x$
		
	\end{algorithmic}
\end{breakablealgorithm}

In Algorithm \ref{Al_FPQAA}, the initial success probability $a$ mentioned in Section \ref{SEC_QAA} is unknown, and we preset a parameter $\delta$ satisfied $\delta^2\leqslant a$ instead. If we denote the final success probability as $P_l=|\langle t|S_l|\psi_0\rangle|^2$ similar to Eq.\eqref{EQ_Pro_Pm}, and denote the initial success probability as $P_0=a$. Then, the following theorem is used to summarize the fixed-point quantum amplitude amplification \cite{RefYoder2014}.
\begin{theorem} \label{TH_FPQAA}
	For any $\delta\in(0,1)$ and $\varepsilon\in(0,1)$, let $\mathcal{A}$ be any quantum algorithm without measurements. Given a Boolean function $f:\{0,1\}^n\rightarrow \{0,1\}$,  if the initial success probability $P_{0}\geqslant \delta^2$,  then the probability that Algorithm \ref{Al_FPQAA} can output an $x$ with $f(x)=1$ is $P_l\geqslant1-\varepsilon^2$ as long as the number of iterations $l\geqslant\lceil\frac{\ln(2/\varepsilon)}{2\delta}\rceil$. 
\end{theorem}

\section{Distributed Quantum Amplitude Amplification}\label{SEC_DQAA}
In this section, we give a theorem to illustrate a relationship between a given Boolean function and its sub-functions, which can be utilized in distributed quantum computing. Then, a distributed quantum amplitude amplification algorithm based on the fixed-point quantum search is proposed.

\subsection{Boolean function and its sub-functions}\label{SEC_B_SB}

As mentioned in Section \ref{SEC_QAA}, a quantum algorithm (unitary operator) $\mathcal{A}$ without measurement is applied to $|0\rangle^{\otimes n}$ in quantum amplitude amplification. To design a distributed quantum algorithm, we assume operator $\mathcal{A}$ can be described as a tensor product of two (or more than two, for example, the tensor product of $n$ Hadamard gates ) unitary sub-operators as follows:
\begin{equation}
	\mathcal{A}=\mathcal{A}_1\otimes\mathcal{A}_2.
	\label{divide_1}
\end{equation} 
Suppose $\mathcal{A}_1$ is an operator acting on the first $j$ qubits, with $0<j<n$. 
Let a Boolean function $f:\{0,1\}^n\rightarrow \{0,1\}$, and it is divided  into $2^j$ sub-functions:
\begin{equation}
	f_k(x)=f(i_k x)
	\label{EQ_SUBF}
\end{equation}
where $0\leq k\leq 2^j-1$, $i_k\in\{0,1\}^j$, $i_k$ is the binary representation of $k$, $x\in\{0,1\}^{n-j}$. 

For any given integer $k$, suppose
\begin{equation}
	|\psi\rangle=\mathcal{A}_2|0\rangle^{\otimes n-j}=|\Psi_{1,k}\rangle+|\Psi_{0,k}\rangle
\end{equation}
where
\begin{align*}
	|\Psi_{1,k}\rangle=\sum_{x:f_k(x)=1}^{}\beta_{x,k}|x\rangle, \\ |\Psi_{0,k}\rangle=\sum_{x:f_k(x)=0}^{}\beta_{x,k}|x\rangle.
\end{align*}
We denote $a_k=\langle\Psi_{1,k}|\Psi_{1,k}\rangle$ and it is thought of as the initial success probability in the node $k$ acted by $\mathcal{A}_2$. 

Then, for each sub-function defined above, similarly to Eq. \eqref{EQ_Q_Generlize}, the distributed iterative operators in distributed quantum amplitude amplification can be represented as
\begin{equation}\label{EQ_Generlize_DQAAO}
	Q_k(\phi,\varphi)=Q(\mathcal{A}_2, f_k, \phi,\varphi)=-\mathcal{A}_2S_0(\phi){\mathcal{A}_2}^{\dagger}S_{f_k}(\varphi) 
\end{equation}
for $0\leq k\leq 2^j-1$, where 
\begin{equation}\label{EQ_Sf_D_Generlize_QAAO}
	S_{f_k}(\varphi)|x\rangle=
	\left\{ 
	\begin{array}{lrc}
		e^{\mathbf{i}\varphi}|x\rangle,&f_k(x)=1,\\
		|x\rangle,&f_k(x)=0,
	\end{array}	
	\right.
\end{equation}
and
\begin{equation}\label{EQ_S0_D_Generlize_QAAO}
	S_0(\phi)|x\rangle=
	\left\{ 
	\begin{array}{lrc}
		e^{-\mathbf{i}\phi}|x\rangle,&x=0^{n-j},\\
		|x\rangle,&x\neq 0^{n-j}.
	\end{array}	
	\right.
\end{equation}

\begin{remark}
	In practice, $j$ is usually a relatively small integer, so $2^j$  can be thought of as a quantity of constant order.
\end{remark}

\begin{remark}
	If $\mathcal{A}$ can be  divided into a more general form as
	\begin{equation}
		\mathcal{A}=\mathcal{A}_1\otimes\mathcal{A}_2\otimes...\otimes\mathcal{A}_t
		\label{divide_2}
	\end{equation} 
	where $t\geqslant 2$, then, as above, we can obtain $2^j$ sub-functions based on the operator $\mathcal{A}_p$ ($p\in\{1,2,...,t\}$) acting on $j$ qubits. The choice of $\mathcal{A}_p$ can be determined by the number of qubits available for computation at each node or by the number of nodes in the computation. For example, in the distributed Grover's algorithms  designed by Qiu et al. \cite{RefQiu2024},   $\mathcal{A}$ is a tensor product of Hadamard  operators, i.e., $\mathcal{A}=H\otimes H\otimes...\otimes H$. 
\end{remark}

Based on the sub-functions in Eq. \eqref{EQ_SUBF}, we can use quantum nodes to compute sub-functions respectively in distributed situation. As is well known, the initial success probability is assumed to be known in the original quantum amplitude amplification  \cite{RefBrassard2002}. However, after dividing the Boolean function $f$ into sub-functions, the initial success probability for each sub-function denoted as $a_k$ becomes unknown. In this case, we can not directly utilize the original quantum amplification to obtain a string $x$ such that $f(x)=1$ when $a_k$ is unknown. 

Fortunately, we discover an essential relationship between $a$ and $a_k$, and this is important for designing our distributed quantum amplitude amplification algorithm. 
We summarize it as follows.

\begin{theorem}\label{TH_range_ak}
	Given a Boolean function $f:\{0,1\}^n\rightarrow \{0,1\}$, if $a$ is the initial probability of $f(x)=1$, then there exists at least one $k\in \{0, 1, ..., 2^j-1\}$ satisfying $a_k\geqslant a$, where $a_k$ is the initial probability of obtaining $x$ with $f_k(x)=1$ for the $k$-th sub-function $f_k(x)$ illustrated in Eq. \eqref{EQ_SUBF}.   
\end{theorem}
\begin{proof}
	Suppose
	\begin{align}
		|\psi\rangle&=(\mathcal{A}_1|0\rangle^{\otimes j})\otimes(\mathcal{A}_2|0\rangle^{\otimes n-j})\\
		&=\sum_{k=0}^{2^j-1}\alpha_{k}|i_k\rangle \sum_{q=0}^{2^{n-j}-1}\beta_{q}|i_q\rangle\\
		&=\sum_{k=0}^{2^j-1}\sum_{q=0}^{2^{n-j}-1}\alpha_{k}\beta_{q}|i_ki_q\rangle \\
		&=\sum_{f(i_ki_q)=1}^{}\alpha_{k}\beta_{q}|i_ki_q\rangle+\sum_{f(i_ki_q)=0}^{}\alpha_{k}\beta_{q}|i_ki_q\rangle,
	\end{align}
	where $k\in \{0, 1, ..., 2^j-1\}$, $q\in \{0, 1, ..., 2^{n-j}-1\}$, $i_k$ and $i_q$ are the binary representation of $k$ and $q$, respectively.
	
	Denote the total probability that target strings are included in $k$-th sub-function as $a[k]$. That is,
	\begin{equation}
		a[k]=\sum_{q:f(i_ki_q)=1}^{}|\alpha_{k}\beta_{q}|^2.
	\end{equation} 

\begin{figure*}[t!]
	\centering
	\includegraphics[scale=0.48]{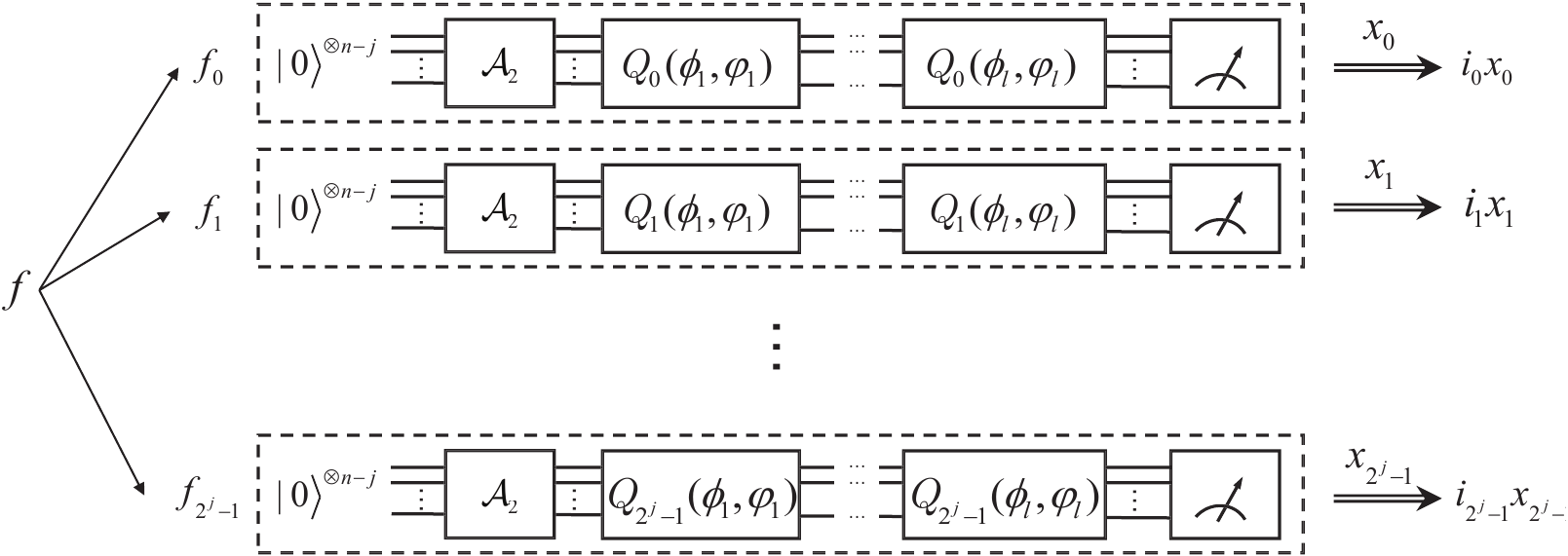}
	\caption{The procedure of distributed quantum amplitude amplification algorithm.}
	\label{FIG_DFPQAA}
\end{figure*}

	Then, we have
	\begin{equation}
		\sum_{k=0}^{2^j-1}a[k]+\sum_{k=0}^{2^j-1}\overline{a[k]}=1
	\end{equation} 
	where $\overline{a[k]}$ is the total probability that non-target strings are included in $k$-th sub-function. Moreover, we have
	\begin{equation}
		a=\sum_{k=0}^{2^j-1}a[k]=\sum_{f(i_ki_q)=1}^{}|\alpha_{k}\beta_{q}|^2
	\end{equation}
	and
	\begin{align}
		a_k&=\langle\Psi_{1,k}|\Psi_{1,k}\rangle\\
		&=\sum\limits_{i_q:f_k(i_q)=1}^{}|\beta_{i_q,k}|^2\\
		&=\frac{\sum\limits_{q:f(i_ki_q)=1}^{}|\alpha_{k}\beta_{q}|^2}{|\alpha_k|^2}\\
		&=\frac{\sum\limits_{q:f(i_ki_q)=1}^{}|\alpha_{k}\beta_{q}|^2}{\sum\limits_{q:f(i_ki_q)=1}^{}|\alpha_{k}\beta_{q}|^2+\sum\limits_{q:f(i_ki_q)=0}^{}|\alpha_{k}\beta_{q}|^2}\\
		&=\frac{a[k]}{a[k]+\overline{a[k]}}.
	\end{align}
	Denote $\dot{k}$ as the $k$-th sub-function satisfying $a[\dot{k}]+\overline{a[\dot{k}]}=0$ and denote $\check{k}$ as the $k$-th sub-function satisfying $a[\check{k}]+\overline{a[\check{k}]}\neq 0$. Therefore, $a[\dot{k}]=0$ and $\overline{a[\dot{k}]}=0$.

	Suppose $a_k<a$ for any $k\in \{0, 1, ..., 2^j-1\}$.  Then we have 
	\begin{equation}
		a[\check{k}]<a(a[\check{k}]+\overline{a[\check{k}]}),
	\end{equation}
	and therefore,
	\begin{align}
		\sum_{k=0}^{2^j-1}a[k]&=\sum a[\dot{k}] + \sum a[\check{k}]\\
		&<\sum a[\dot{k}]+ \sum a(a[\check{k}]+\overline{a[\check{k}]})\\
		&=\sum a(a[\dot{k}]+\overline{a[\dot{k}]})+\sum a(a[\check{k}]+\overline{a[\check{k}]})\\
		&=\sum_{k=0}^{2^j-1}(a(a[k]+\overline{a[k]}))\\
		&=a(\sum_{k=0}^{2^j-1}a[k]+\sum_{k=0}^{2^j-1}\overline{a[k]})\\
		&=a.
	\end{align}
	So, the assumption leads to contradiction, and  the proof  is completed.
\end{proof}

\subsection{Distributed Quantum Amplitude Amplification algorithm}\label{FPDQAA}
With Theorem \ref{TH_range_ak} in  Section \ref{SEC_B_SB},  we can design a distributed quantum amplitude amplification algorithm.

According to Section \ref{Sec_FPQAA}, if we can find a lower bound of the initial success probability, then we can utilize the fixed-point quantum search to find a target string within arbitrary probability $1-\varepsilon^2$, where $\varepsilon$ is a preset parameter. Therefore, by combining Theorem \ref{TH_FPQAA} with Theorem \ref{TH_range_ak}, we can set the initial success probability of each quantum node as $a$ to obtain at least one target string from a quantum node. The number of iterations $l$ is then determined by $a$ and $\varepsilon$. Consequently, the iterative operators in the node $k$ are defined as
\begin{align}\label{EQ_Q_D_Generlize}
	Q_k(\phi_r,\varphi_r)&=Q(\mathcal{A}_2, f_k, \phi_r,\varphi_r)
\end{align}
for $r=1, 2, ..., l$, where $\phi_r$ and $\varphi_r$ in the operators are defined as Eq. \eqref{phi_varphi_r}.

Then, we can give a distributed quantum amplitude amplification algorithm as follows.

\begin{breakablealgorithm}
	\caption{Distributed quantum amplitude amplification algorithm}
	\label{AL_DQAFPQS}
	\begin{algorithmic}[1]
		\REQUIRE The iterative operators of each node Algorithm $\mathcal{A}$,  function $f$,  error rate $\varepsilon^2$.
		\ENSURE $x\in \{0,1\}^{n}$ such that $f(x)=1$.
		\STATE Given $j\in[1,n)$, divide $f$ into $f_k (k=0, 1, ... , 2^{j}-1)$ as Eq. \eqref{EQ_SUBF}.
		\STATE $\mathbf{All~ quantum~ nodes~ perform~ the~ }$\\$\mathbf{computation~ of~ f_k~ according~ to~ the~}$\\$\mathbf{following~ procedure~ in~ parallel.}$
		\STATE Set $l=\left\lceil\frac{\ln(2/\varepsilon)}{2\sqrt{a}}\right\rceil$.
		\STATE $|\psi_0\rangle=\mathcal{A}_2|0\rangle^{\otimes n-j}$.
		\STATE $|\psi_{l,k}\rangle=S_{l,k}|\psi_0\rangle$, where $S_{l,k}=Q_k(\phi_l,\varphi_l)...Q_k(\phi_2,\varphi_2)Q_k(\phi_1,\varphi_1)$.
		\STATE Measure the register and obtain $x_k$.
		\STATE Output $x=i_kx_{k}$ satisfying $f(x)=1$, where $ i_k\in\{0,1\}^{j}$ is the binary representation of $k$. 
		\RETURN $x$.
		
	\end{algorithmic}
\end{breakablealgorithm}
\begin{remark}
	If the number of quantum nodes is limited, we can compute $f_k(k=0,1,...,2^j-1)$ sequentially on a single quantum node as \cite{RefQiu2024}.
\end{remark}

The procedure of  algorithm \ref{AL_DQAFPQS} can be depicted by Figure \ref{FIG_DFPQAA}.

Combining Theorem \ref{TH_FPQAA} with Theorem \ref{TH_range_ak}, we give the following theorem to support our distributed quantum amplitude amplification algorithm.

\begin{theorem} \label{TH_DFPQAA}
	Given $n>j\geqslant1$, for any $\varepsilon\in(0,1)$, let $\mathcal{A}$ be any quantum algorithm that uses no measurements, and let a Boolean function $f:\{0,1\}^n\rightarrow \{0,1\}$. If the initial success probability is $a$,
	then there exists at least one $k\in \{0, 1, ..., 2^j-1\}$ satisfying  $a_k\geqslant a$ as Theorem \ref{TH_range_ak}, and the probability that Algorithm \ref{AL_DQAFPQS} can output an $x$ such that $f(x)=1$ is at least $1-\varepsilon^2$ for the number of iterations $l\geqslant\lceil\frac{\ln(2/\varepsilon)}{2\sqrt{a}}\rceil$. 
\end{theorem}
\begin{proof}
	Theorem \ref{TH_range_ak} reveals that at least one quantum node in the algorithm has an initial successful probability exceeding $a$. Denote an arbitrary $k$-th quantum node satisfying this condition as $\hat{k}$ and sequentially denote the indices of all nodes satisfying the condition by $\hat{k_0}$, $\hat{k_1}$, ..., $\hat{k_t}$ for $0\leqslant t\leqslant 2^j-1$. Then, according to Theorem \ref{TH_FPQAA}, the final success probability of $\hat{k}$-th quantum node $P_{\hat{k}}	\geqslant 1-\varepsilon^2$ as long as the number of iterations is not less than $\lceil\frac{\ln(2/\varepsilon)}{2\sqrt{a_{\hat{k}}}}\rceil$.
	
	As a result, after $l\geqslant\lceil\frac{\ln(2/\varepsilon)}{2\sqrt{a}}\rceil\geqslant\lceil\frac{\ln(2/\varepsilon)}{2\sqrt{a_{\hat{k}}}}\rceil$ iterations, we have
	\begin{align}
		\nonumber
		P&=1-(1-P_{\hat{k_0}})(1-P_{\hat{k_1}})...(1-P_{\hat{k_t}})\\\nonumber
		&\geqslant P_{\hat{k}}\\
		&\geqslant 1-\varepsilon^2.
	\end{align}
	The proof of theorem is completed.
\end{proof}

In the distributed quantum amplitude amplification algorithm above, the query complexity depends on the initial success probability parameter $a$ and the preset final error rate $\varepsilon^2$. According to Theorem \ref{TH_DFPQAA}, the query complexity of algorithm \ref{AL_DQAFPQS} is $l=O(\frac{\log(2/\varepsilon)}{2\sqrt{a}})$.

\subsection{Simulation and Comparisons}\label{SEC_SIM_COM}
In this subsection, to further validate our algorithm, we conduct simulations of the proposed distributed algorithm via Qiskit. After that, comparisons between our algorithm and the related works are given.

\subsubsection{Simulation for Algorithm \ref{AL_DQAFPQS}}\label{SUB_SEC_SIMDDFPQAA}
In this subsection, we first present the circuit utilized in the simulation and give our simulation results in the case of Grover's algorithm. 
\begin{figure}[!t]
	\centering
	\includegraphics[scale=0.4]{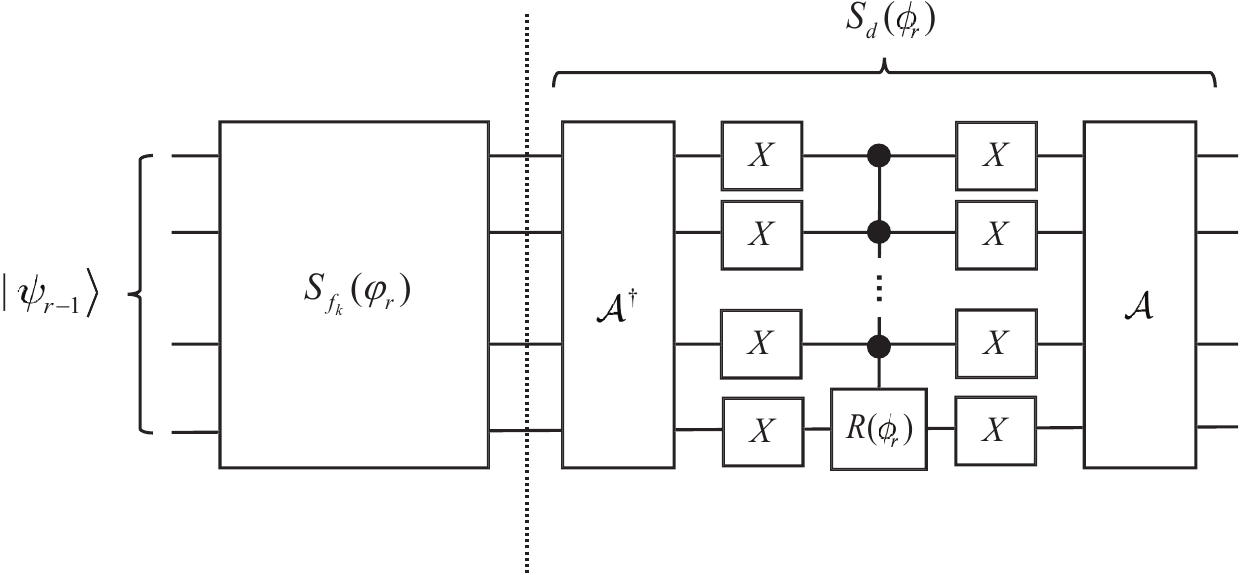}
	\caption{The circuit of iterative operators $Q_k(\phi_r, \varphi_r)$ in Algorithm \ref{AL_DQAFPQS}.}
	\label{FIG_cicuit}
\end{figure}
Depicted in the Figure \ref{FIG_cicuit}, the circuit of an iteration in Algorithm \ref{AL_DQAFPQS} can be divided into two parts: the oracle $S_f(\varphi_r)$ and the diffusion operator $S_d(\phi_r)=\mathcal{A}S_0(\phi_r)\mathcal{A}^{\dagger}$. $R(\phi_r)$ in the Figure \ref{FIG_cicuit} is a phase gate, and
\begin{equation*}
	R(\phi_r)=\left(\begin{array}{cc}
		1&0\\
		0&e^{-\mathbf{i}\phi_r}
	\end{array}\right).
\end{equation*}

Grover's algorithm is a special quantum amplitude amplification algorithm with $\mathcal{A}=H^{\otimes n}$, where $n$ is the number of qubits.

Let a Boolean function $f:\{0,1\}^6\rightarrow\{0,1\}$. Suppose these elements $x$ in $\{110110,111111,011001\}$ are target strings such that $f(x)=1$.
Set $\varepsilon=0.3$ and $\mathcal{A}_1=H^{\otimes 2}$. Then we repeat the simulation $1000$ times and obtain the following outputs of each quantum node.

\begin{figure*}[t!]
	\centering
	
	\subfloat[Result of the node 0.]{%
		\includegraphics[width=0.35\textwidth]{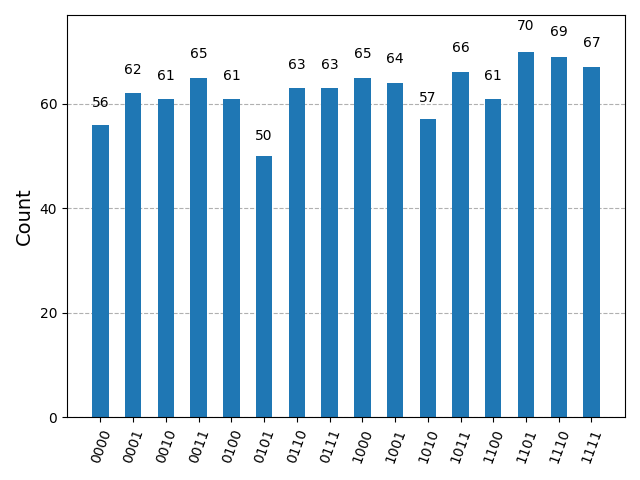}
		\label{sFG_Result_DGA0}
	}
	\quad
	\subfloat[Result of the node 1.]{
		\includegraphics[width=0.35\textwidth]{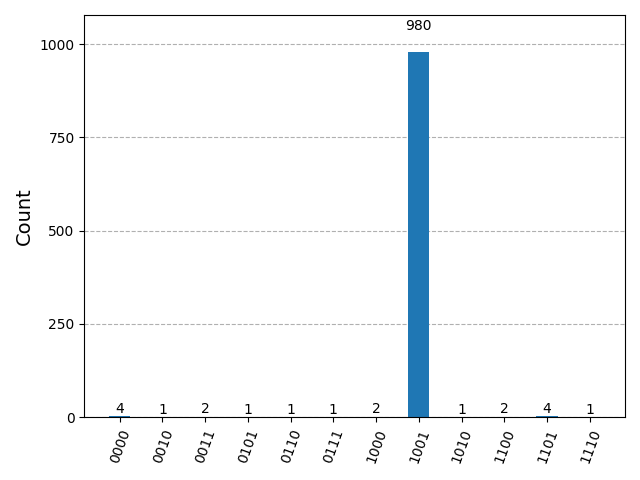}
		\label{sFG_Result_DGA1}
	}
	
	\vspace{1em}
	
	\subfloat[Result of the node 2.]{
		\includegraphics[width=0.35\textwidth]{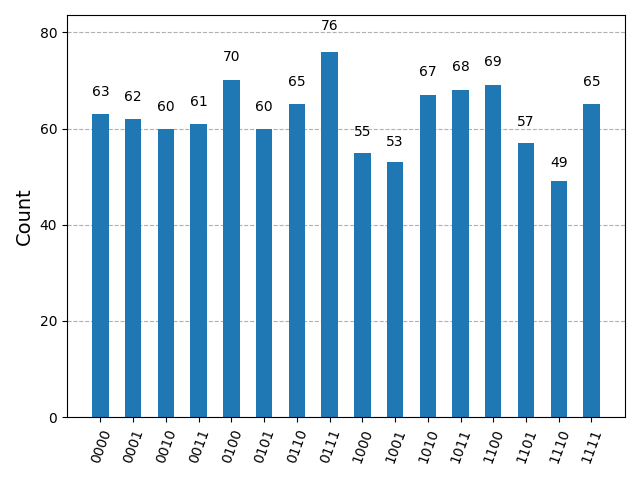}
		\label{sFG_Result_DGA2}
	}
	\quad
	\subfloat[Result of the node 3.]{
		\includegraphics[width=0.35\textwidth]{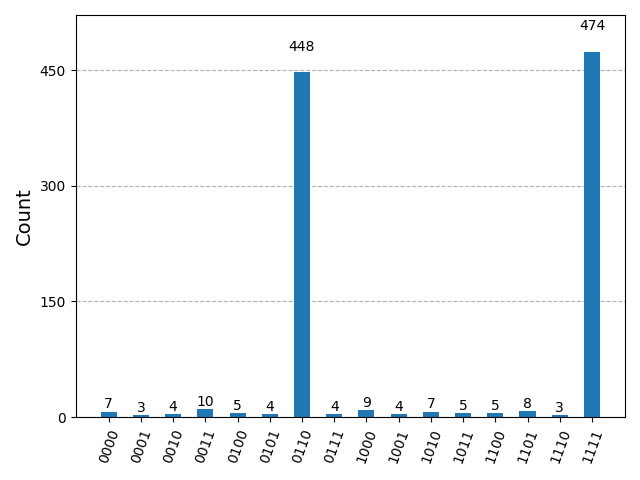}
		\label{sFG_Result_DGA3}
	}
	
	\caption{Result of Algorithm \ref{AL_DQAFPQS} in each quantum node with $\varepsilon=0.3$ and $\mathcal{A}_1=H^{\otimes 2}$. The target strings are $\{110110, 111111, 011001\}$. Because target strings are not included in the node $0$ and the node $2$, the probabilities of output are random as Figure \ref{sFG_Result_DGA0} and \ref{sFG_Result_DGA2}. Meanwhile, the node $1$ and the node $3$ can output target strings with high probability. Additionally, it means that after running the algorithm once, we can obtain two target strings simultaneously with a high probability.}
	\label{FG_Result_DGA}
\end{figure*}

According to the results, the quantum node without target strings have random outputs as shown in Figure \ref{sFG_Result_DGA0} and Figure \ref{sFG_Result_DGA2}. Meanwhile, The results in Figure \ref{sFG_Result_DGA1} and Figure \ref{sFG_Result_DGA3} show that the corresponding nodes output target strings with probability more than $1-\varepsilon^2=0.91$ due to both the practical initial success probability $a_k$ exceeding the whole initial success probability $a$. Therefore, the simulation corroborates the correctness of the proposed algorithm.

\subsubsection{Comparisons to the Original Quantum Amplitude Amplification}
In this subsection, we compare Algorithm \ref{AL_DQAFPQS} with the original  quantum amplitude amplification algorithm by Brassard et al.  \cite{RefBrassard2002}.

\begin{figure*}[t!]
	\centering
	
	\subfloat[Result of the Algorithm \ref{AL_DQAFPQS}.]{%
		\includegraphics[width=1\textwidth]{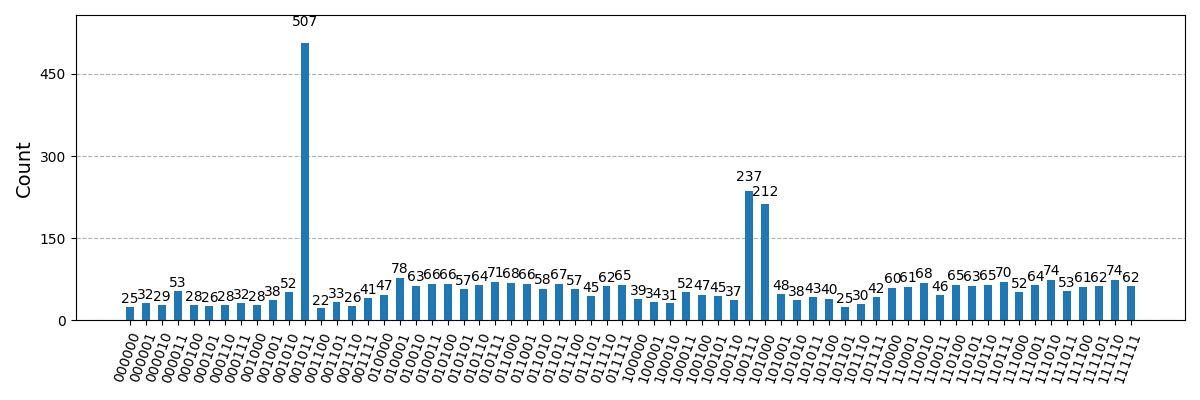}
		\label{noiseDFPQAA}
	}

	\vspace{1em}
	
	\subfloat[Result of the Algorithm \ref{QAA_a1}, and $\mathcal{A}=H^{\otimes 6}$]{%
		\includegraphics[width=1\textwidth]{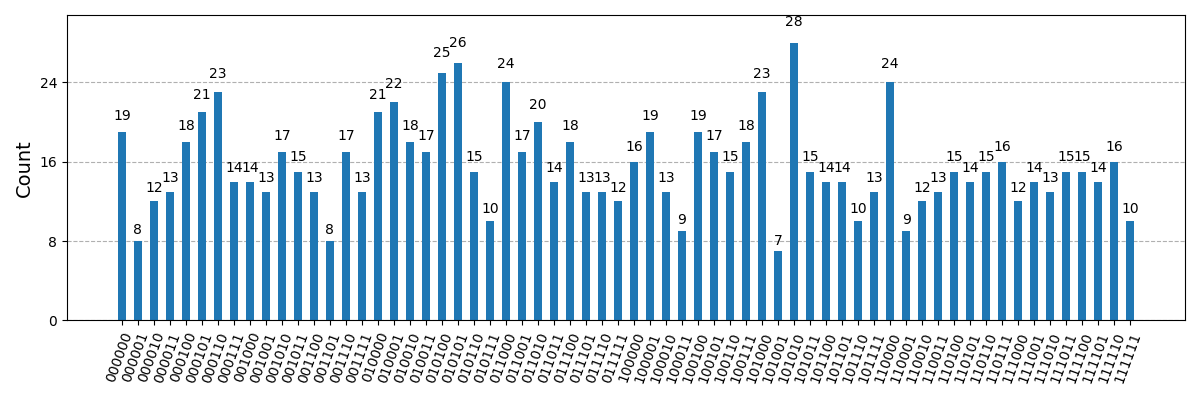}
		\label{noiseGrover}
	}

	\caption{Compared the performance of the proposed algorithm with the original quantum amplitude amplification \cite{RefBrassard2002} in noise environment. The noisy simulator we use is GenericBackendV2 of Qiskit. Additionally, $\varepsilon=0.65$, $\mathcal{A}_1=H^{\otimes 2}$ and the target states are $\{110110,111111,011001\}$. We repeatedly execute algorithms for $1000$ times. Figure \ref{noiseDFPQAA} provides an overall view of the output distributions from all four quantum nodes in the presence of noise (hence the total number of counts in the figure is 4000). Figure \ref{noiseGrover} shows the output of Algorithm \ref{QAA_a1} when searching for the same target under noisy conditions. It is clear that, when searching for a 6-bit string in a noisy setting, our algorithm can still find the target string with a high probability(even more than one target string), while Algorithm \ref{QAA_a1} can no longer do so.}
	\label{FG_Result_noise}
\end{figure*}

According to Section \ref{FPDQAA}, Algorithm \ref{AL_DQAFPQS} reduces the required qubits from $n$ to $n-j$. It also means that it cost less single-qubit gates and CNOT gates to construct an iterative operator. This advantage enables our algorithm to perform better even in noisy environments as Figure \ref{FG_Result_noise}.

In Figure \ref{FG_Result_noise}, since the parameter $\varepsilon$ in Algorithm \ref{AL_DQAFPQS}  
is preset, we set $\varepsilon=0.65$ to compare the two algorithms under the same query complexity (or even under a lower query complexity for our algorithm), in other words, $l=\left\lceil\frac{\ln(2/\varepsilon)}{2\sqrt{a}}\right\rceil\leqslant\lfloor\frac{\pi}{4\arcsin(\sqrt{a})}\rfloor$. Here, $a=\frac{3}{64}$. After that, it is obvious  to find that our algorithm performs better in the noise environment according to the simulation results.

\subsubsection{Comparisons to Other Related Works}

Furthermore, we compare Algorithm \ref{AL_DQAFPQS} with  other related works in this subsection.
On one hand, we compare Algorithm \ref{AL_DQAFPQS} with the parallel distributed Grover's algorithm proposed by Qiu et al. \cite{RefQiu2024}. Alternatively, distributed quantum amplitude amplification can also be achieved by using quantum amplitude estimation \cite{RefBrassard2002}. Specifically, Theorem \ref{TH_range_ak} provides a range for $\widetilde{a_k}$ (the estimation of $a_k$) that allows us to determine the bound on the number of iterations. Subsequently, by repeatedly executing the quantum amplitude amplification algorithm within this iteration boundary of each quantum node, target states can be identified, as demonstrated in \cite{RefQiu2024}. In a way, our approach has generalized the parallel distributed Grover's algorithm in \cite{RefQiu2024}.

On the other hand,  quantum communication is usually required in other distributed quantum algorithms for search problems or amplitude amplification (e.g., \cite{RefLiH2024G}), but in Algorithm \ref{AL_DQAFPQS} , no quantum communication is needed.

\section{More Applications to Distributed Quantum Computing}\label{SEC_APPL_QDC}
In this section, we consider more applications of the fixed-point quantum search algorithm proposed by Yoder et al. \cite{RefYoder2014} to distributed quantum computing. Firstly, we give a more general model inspired by Algorithm \ref{AL_DQAFPQS}. Then, we present some design strategies of distributed quantum algorithms to further illustrate specific applications in three classes of problems: the SAT problem, the element distinctness problem, and the graph problems.

\subsection{General Model}\label{SEC_GM}
Before proposing the more general model, we firstly review the design strategy of the distributed quantum amplitude amplification algorithms described in Section \ref{SEC_DQAA}. 

A large-scale search problem or amplitude amplification problem is decomposed into multiple sub-problems, which are then processed in parallel by different quantum nodes. Such a distributed decomposition generally makes the initial proportion of solutions in each sub-problem unknown, and the fixed-point quantum search can address this issue to a certain extent. Specifically, according to Theorem \ref{TH_FPQAA}, if these sub-problems may be either solvable or unsolvable, but their initial success probability is at least $\delta^2$ whenever a solution exists, then the fixed-point quantum search can be used to find a solution or to return information indicating that no solution exists with a prescribed success probability $1-\varepsilon^2$. Finally, the solutions to these sub-problems are recombined through classical post-processing to construct a solution to the original problem. Therefore, a more general model is given as follows.

\begin{breakablealgorithm}
	\caption{A general model for distributed quantum algorithms based on fixed-point quantum search algorithm}
	\label{AL_FPforDQC}
	\begin{algorithmic}[1]
		\REQUIRE The problem $P$, the number of nodes $M$, parameter $\delta$, error rate $\varepsilon^2$.
		\ENSURE The solution of the problem.
		\STATE Divide the problem into $M$ sub-problems and assign each sub-problem to one quantum node for parallel computation.
		\STATE Each quantum node performs the fixed-point quantum search with parameter $\delta^2$.
		\STATE Verify the output of each node. Retain the solutions verified as correct at each node and discard the incorrect ones
		\STATE Combine the correct solutions obtained by the individual nodes classically to construct a solution to the complete problem.
		\RETURN The solution of the problem.		
	\end{algorithmic}
\end{breakablealgorithm}
Under the assumptions stated above, if the correctness of the final result depends on the success probability of the output from all nodes, then the probability of obtaining a correct final result under the proposed model is
\begin{align}
	\nonumber
	P&=P_{0}P_{1}...P_{M}\\
	&\geqslant (1-\varepsilon^2)^M.\label{EQ_P_1}
\end{align}
In other cases, the probability of obtaining a correct final result is also usually higher than $(1-\varepsilon^2)^M$. For example, the success probability of Algorithm \ref{AL_DQAFPQS} is at least $1-\varepsilon^2$ as shown in Theorem \ref{TH_DFPQAA}.
In particular, if the output of any individual node can be used to recover the correct final result, then the success probability under this model is 
\begin{align}
	\nonumber
	P&=1-(1-P_{0})(1-P_{1})...(1-P_{M})\\\nonumber
	&\geqslant 1-(1-(1-\varepsilon^2))^M\\
	&\geqslant 1-(\varepsilon^{2M}).\label{EQ_P_2}
\end{align}
Moreover, the distributed quantum algorithm implementing this model requires fewer quantum resources at each node and involves no additional quantum communication.
%

In the rest of this section, we focus on three classes of problems (the SAT problem, the element distinctness problem, and the graph problems) to illustrate some special applications of the fixed-point quantum search to distributed quantum computing with above model. Besides some strategies given to design distributed quantum algorithms for these problems, we also show a general strategy for the setting of the  fixed-point quantum search to handle the sub-problems. Specifically, if the input to the quantum algorithm is a uniform superposition state, then the parameter $\delta$ can be set as $\frac{1}{\sqrt{N}}$, where $N$ is the size of search space in each node.

\subsection{SAT Problem}\label{SEC_SAT}
The Boolean satisfiability (SAT) problem \cite{RefCook1971} is a well-known NP-complete problem in computer science. For any given Boolean function $f:\{0,1\}^n\rightarrow \{0,1\}$, the $n$-SAT problem is to determine whether there exists a string $x=x_1x_2...x_n$ consisting of $n$ input variables $x_1, x_2, ..., x_n$ such that $f(x)=1$.

The Boolean function $f:\{0,1\}^n\rightarrow \{0,1\}$ in $n$-SAT problem is always given as a CNF with $M$ clauses, that is, $f$ can be represented as 
\begin{equation}
	f=C_1\wedge C_2\wedge...\wedge C_M
\end{equation}
where $C_k$ ($k\in\{1,...,M\}$) is a clause formed by disjunction of variables or  negation of these variables from $\{x_1, x_2, ..., x_n \}$. For example, $x_1\vee x_3$, $\neg x_2\vee x_3$ are clauses and $(x_1\vee x_3)\wedge(\neg x_2\vee x_3)$ is a CNF with two clauses.

If the Boolean function for the $n$-SAT problem is given in CNF, the most direct distributed approach is to assign each clause of the CNF formula to an individual node for computation. This can reduce the size of the oracle and the time complexity. Since the number of solutions satisfying each clause is unknown after the distribution, the fixed-point quantum search can be employed to solve these sub-problems, with the parameter $\delta^2=\frac{1}{2^n}$. In this case, because the final result is related to the output of all nodes, the success probability is at least $(1-\varepsilon^2)^M$ similar to Eq. \eqref{EQ_P_1}.

\subsection{Element Distinctness Problem}\label{SEC_ED}
The element distinctness problem is defined as follows.
\begin{definition}
	Given two finite sets $S_1$, $S_2$ and a function $f:S_1\rightarrow S_2$, our target is to determine whether there exist two distinct elements $i, j\in S_1$ such that $f(i)=f(j)$. Moreover, if such a pair exists, we need to identify any pair $i,j$ satisfying the condition.
\end{definition}

In 2004, Ambainis \cite{RefAmbainis2004} reformulated the quantum algorithm proposed by Buhrman et al. \cite{RefBuhrman2005} for solving the element distinctness problem. The procedure is as follows.\\
\textbf{Step 1:} Choose $\lceil\sqrt{|S_1|}\rceil$ random elements in $S_1$ and sequentially denote the indices of chosen elements as $e_1, ..., e_{\lceil\sqrt{|S_1|}\rceil}$. In other words, $e_1,$ $ ..., e_{\lceil\sqrt{|S_1|}\rceil} \in S_1$. Evaluate $f(e_1), ..., f(e_{\lceil\sqrt{|S_1|}\rceil})$. If two of them are equal, stop, output the two equal elements.
\\
\textbf{Step 2:} Use Grover's algorithm to search among remaining $|S_1|-\lceil\sqrt{|S_1|}\rceil$ elements in $S_1$ for an element $\hat{e}$ such that $f(\hat{e})=f(e_k)$ for some $k\in\{1, 2, ..., \lceil\sqrt{|S_1|}\rceil\}$.

We consider to replace \textbf{Step 2} above with a distributed quantum algorithm under the model described in Section \ref{SEC_GM}. The procedure is as follows.
\\
\textbf{Step 2 (Modified):} Given $M$ quantum nodes, divide the remaining $|S_1|-\lceil\sqrt{|S_1|}\rceil$ elements as evenly as possible into $M$ subsets ${S_1}^{(1)}, {S_1}^{(2)}, ..., {S_1}^{(M)}$. All $M$ quantum nodes perform the following procedure in parallel. For each $r\in\{1, 2, ..., M\}$, the $r$-th quantum node uses the fixed-point quantum search algorithm to search among at most  $\lceil\frac{|S_1|-\lceil\sqrt{|S_1|}\rceil}{M}\rceil$ elements in ${S_1}^{(r)}$ for an element $\hat{e_r}$ such that $f(\hat{e_r})=f(e_k)$ with some $k\in\{1, 2, ..., \lceil\sqrt{|S_1|}\rceil\}$.

If there exists a pair $i$, $j$ such that $f(i)=f(j)$, then the probability that $i$ is one of the randomly selected elements $e_1, ..., e_{\lceil\sqrt{|S_1|}\rceil}$ is $\frac{\lceil\sqrt{|S_1|}\rceil}{|S_1|}$. Therefore, the amplitude amplification is applied to the algorithm proposed by Buhrman et al. to increase the success probability in \cite{RefAmbainis2004}.  

However, our optimization only focuses on \textbf{Step 2}. \textbf{Step 2 (Modified)} can reduce the size of the search space computed by each quantum node. Meanwhile, because we only need to determine at least one collision pair, \textbf{Step 2 (Modified)} may further increase the success probability of \textbf{Step 2} in \cite{RefBuhrman2005} to $1-\varepsilon^2$, which is similar to that of Algorithm \ref{AL_DQAFPQS} illustrated in Theorem \ref{TH_DFPQAA}. Since the number of potential collision elements in the subsets is unknown, the parameter of the fixed-point quantum search algorithm of \textbf{Step 2 (Modified)} can be set as $\delta_r=\frac{1}{\sqrt{|{S_1}^{(r)}|}}$ for $r\in\{1, 2, ..., M\}$.

\subsection{Graph Problems}\label{SEC_GP}
For graph-related problems, such as graph coloring and the Hamiltonian cycle problem \cite{RefKarp1972}, relevant quantum algorithms usually involve Grover's algorithm or the amplitude amplification algorithm. 

Consider a search problem defined on a graph $G=(V,E)$, where $V$ is the vertex set and $E$ is the edge set. A solution to the problem can usually be represented by a string $x$ encoding information about the edges or vertices. If the graph $G$ can be decomposed into sub-graphs $G_1, G_2, ..., G_M$, and the corresponding sub-problems can be formulated by these sub-graphs, then we may have a distributed quantum algorithm as follows. First, we use $M$ quantum nodes to solve the problem defined on the corresponding sub-graph in parallel. Then, we will obtain some sub-strings $x_1, x_2, ..., x_M$. Finally, these sub-strings are combined through splicing or other classical post-processing procedures to construct a solution $x$ to the original problem.

In the above procedure, since the number of target solutions for each sub-graph search problem is unknown, the fixed-point quantum search can be employed to solve these problems. The parameter of the fixed-point search  for each sub-graph can usually be set as $\delta=\frac{1}{\sqrt{2^{|x_k|}}}$ for $k\in\{1, 2 , ..., M\}$, where $|x_k|$ denotes the length of the binary string $x_k$. Because the correctness of the final solution is related to the combination of the correct solutions for all sub-graphs, the success probability of this case is at least  $(1-\varepsilon^2)^M$ as shown in Eq. \eqref{EQ_P_1}.

\section{Conclusions}\label{SEC_Con}
In this paper, we  first revisited the fixed-point quantum search and observed that its properties can be exploited in the design of distributed quantum algorithms. Then, we have proposed a distributed quantum amplitude amplification algorithm based on the fixed-point quantum search with certain advantages: (1)  the number of qubits required for each quantum node is reduced; (2)  the quantum gates required in the iterative operators are decreased; (3) neither quantum communication nor quantum counting is required; (4) the robustness under the influence of noise is satisfactory.  Furthermore, we have simulated our algorithm by means of Qiskit and discussed how our distributed quantum algorithm  has  generalized the distributed Grover's algorithm in \cite{RefQiu2024} to an extent. 

Inspired by the proposed algorithm, we have formulated a general model for applying the fixed-point quantum search to distributed quantum computing and then presented several specific application scenarios. This model may provide certain insights and ideas to guide the design of distributed quantum algorithms for other specific problems.

Theorem \ref{TH_range_ak}  reveals an inherent relationship between any given Boolean function and its sub-functions and may be helpful for designing other distributed algorithms or related algorithms. Therefore, an interesting question for future study is to discover more applications of Theorem \ref{TH_range_ak}.

Furthermore, the optimization problems are one of the important research topics in the applications of quantum algorithms. Grover Adaptive Search (GAS) \cite{RefGilliam2021,RefBulger2003,RefBaritompa2005} has been used to transform an optimization problem into a sequence of search problems with dynamically updated oracles. In a way, the fixed-point quantum search may replace another interesting search \cite{RefBoyer1998} to solve the overcooking problem in the quantum search procedure of GAS. So, another problem worthy of consideration is to investigate distributed GAS algorithms based on the fixed-point quantum search, and we would like to study it in future.


\begin{thebibliography}{00}


\bibitem{RefKnorzer2026}
J. Kn{\"o}rzer, X. Liu, B. F. Schiffer, and J. Tura, ``Distributed quantum information processing: a review of recent progress,''
\href{https://doi.org/10.1088/1361-6633/ae74e0}
{\emph{Reports on Progress in Physics} \textbf{89}, 074401 (2026)}.

\bibitem{RefQiu2024}
D. W. Qiu, L. Luo, and L. G. Xiao, ``Distributed Grover's algorithm,''
\href{https://doi.org/10.1016/j.tcs.2024.114461}
{\emph{Theoretical Computer Science} \textbf{993}, 114461 (2024)}.

\bibitem{RefBeals2013}
R. Beals, S. Brierley, O. Gray, A. W. Harrow, S. Kutin, N. Linden, D. Shepherd, and M. Stather, ``Efficient distributed quantum computing,''
\href{https://doi.org/10.1098/rspa.2012.0686}
{\emph{Proceedings of the Royal Society A: Mathematical, Physical and Engineering Sciences} \textbf{469}, 20120686 (2013)}.

\bibitem{RefLeGall2019}
F. Le Gall, H. Nishimura, and A. Rosmanis, ``Quantum advantage for the LOCAL model in distributed computing,''
\href{https://doi.org/10.4230/LIPIcs.STACS.2019.49}
{\emph{Leibniz International Proceedings in Informatics} \textbf{126}, 49:1--49:14 (2019)}.

\bibitem{RefLi2017}
K. Li, D. W. Qiu, L. Z. Li, S. G. Zheng, and Z. B. Rong, ``Application of distributed semi-quantum computing model in phase estimation,''
\href{https://doi.org/10.1016/j.ipl.2016.12.002}
{\emph{Information Processing Letters} \textbf{120}, 23--29 (2017)}.


\bibitem{RefQiu2025}
D. Qiu, L. Xiao, L. Luo, and P. Mateus, ``Error correction for distributed quantum computing,''
\href{https://doi.org/10.1140/epjqt/s40507-025-00455-x}
{\emph{EPJ Quantum Technology} \textbf{12}, 142 (2025)}.

\bibitem{RefAvron2021}
J. Avron, O. Casper, and I. Rozen, ``Quantum advantage and noise reduction in distributed quantum computing,''
\href{https://doi.org/10.1103/PhysRevA.104.052404}
{\emph{Physical Review A} \textbf{104}, 052404 (2021)}.


\bibitem{RefTan2022}
J. W. Tan, L. G. Xiao, D. W. Qiu, L. Luo, and P. Mateus, ``Distributed quantum algorithm for Simon's problem,''
\href{https://doi.org/10.1103/PhysRevA.106.032417}
{\emph{Physical Review A} \textbf{106}, 032417 (2022)}.

\bibitem{RefLi2024GS}
H. Li, D. W. Qiu, L. Luo, and P. Mateus, ``Exact distributed quantum algorithm for generalized Simon's problem,''
\href{https://doi.org/10.1007/s00236-024-00455-x}
{\emph{Acta Informatica} \textbf{61}, 131--159 (2024)}.

\bibitem{RefYimsiriwattana2004}
A. Yimsiriwattana and S. J. Lomonaco Jr., ``Distributed quantum computing: a distributed Shor algorithm,''
\href{https://doi.org/10.1117/12.546504}
{\emph{Quantum Information and Computation II} \textbf{5436}, 360--372 (2004)}.

\bibitem{RefXiao2023A}
L. G. Xiao, D. W. Qiu, L. Luo, and P. Mateus, ``Distributed Shor's algorithm,''
\href{https://doi.org/10.26421/QIC23.1-2-3}
{\emph{Quantum Information and Computation} \textbf{23}, 27--44 (2023)}.

\bibitem{RefLiH2025}
H. Li, D. Qiu, and L. Luo, ``Distributed Deutsch--Jozsa algorithm,''
\href{https://doi.org/10.1007/s11227-025-07683-z}
{\emph{The Journal of Supercomputing} \textbf{81}, 1221 (2025)}.

\bibitem{RefLiH2024DJ}
H. Li, D. W. Qiu, and L. Luo, ``Distributed generalized Deutsch--Jozsa algorithm,''
\href{https://doi.org/10.1007/978-981-96-1093-8_18}
{\emph{in Computing and Combinatorics: 30th International Conference, COCOON 2024}, \textbf{15162}, 214-225. Springer (2024).}

\bibitem{RefGrover1997}
L. K. Grover, ``Quantum mechanics helps in searching for a needle in a haystack,''
\href{https://doi.org/10.1103/PhysRevLett.79.325}
{\emph{Physical Review Letters} \textbf{79}, 325-328 (1997)}.

\bibitem{RefBrassard2002}
G. Brassard, P. H{\o}yer, M. Mosca, and A. Tapp, ``Quantum amplitude amplification and estimation,''
\href{https://doi.org/10.1090/conm/305/05215}
{\emph{Contemporary Mathematics} \textbf{305}, 53-74 (2002)}.

\bibitem{RefUtsumi2026}
T. Utsumi and Y. Nakata, ``Explicit decoders using fixed-point amplitude amplification based on QSVT,''
\href{https://doi.org/10.22331/q-2026-03-13-2024}
{\emph{Quantum} \textbf{10}, 2024 (2026)}.

\bibitem{RefGilliam2021}
A. Gilliam, S. Woerner, and C. Gonciulea, ``Grover adaptive search for constrained polynomial binary optimization,''
\href{https://doi.org/10.22331/q-2021-04-08-428}
{\emph{Quantum} \textbf{5}, 428 (2021)}.

\bibitem{RefLiH2024G}
H. Li and D. Qiu,
``Distributed multi-objective quantum search algorithm,''
\href{https://doi.org/10.1007/978-981-96-9894-3_36}
{\emph{in Proceedings of the 21st Advanced Intelligent Computing Technology and Applications},
Lecture Notes in Computer Science, vol. 15853, 438-449. Springer (2025).}

\bibitem{RefGrover2005}
L. K. Grover,
``Fixed-point quantum search,''
\href{https://doi.org/10.1103/PhysRevLett.95.150501}
{\emph{Physical Review Letters} \textbf{95}, 150501 (2005)}.

\bibitem{RefYoder2014}
T. J. Yoder, G. H. Low, and I. L. Chuang,
``Fixed-point quantum search with an optimal number of queries,''
\href{https://doi.org/10.1103/PhysRevLett.113.210501}
{\emph{Physical Review Letters} \textbf{113}, 210501 (2014)}.

\bibitem{RefLong1999}
G. L. Long, Y. S. Li, W. L. Zhang, and L. Niu,
``Phase matching in quantum searching,''
\href{https://doi.org/10.1016/S0375-9601(99)00631-3}
{\emph{Physics Letters A} \textbf{262}, 27-34 (1999)}.

\bibitem{RefHoyer2000}
P. H{\o}yer,
``Arbitrary phases in quantum amplitude amplification,''
\href{https://doi.org/10.1103/PhysRevA.62.052304}
{\emph{Physical Review A} \textbf{62}, 052304 (2000)}.

\bibitem{RefRivlin2020}
T. J. Rivlin,
\emph{Chebyshev Polynomials: From Approximation Theory to Algebra and Number Theory},
2nd ed., Dover Publications (2020).

\bibitem{RefCook1971}
S. A. Cook, ``The complexity of theorem-proving procedures,'' 
\href{https://doi.org/10.1145/800157.805047}
{in \emph{Proceedings of the Third Annual ACM Symposium on Theory of Computing}, 151-158, ACM Press (1971)}.

\bibitem{RefAmbainis2004}
A. Ambainis, ``Quantum search algorithms,''
\href{https://doi.org/10.1145/992287.992296}
{\emph{ACM SIGACT News} \textbf{35}, 22--35 (2004)}.

\bibitem{RefBuhrman2005}
H. Buhrman, C. D{\"u}rr, M. Heiligman, P. H{\o}yer, F. Magniez, M. Santha, and R. de Wolf, ``Quantum algorithms for element distinctness,''
\href{https://doi.org/10.1137/S0097539702402780}
{\emph{SIAM Journal on Computing} \textbf{34}, 1324-1330 (2005)}.

\bibitem{RefKarp1972}
R. M. Karp, ``Reducibility among combinatorial problems,'' 
\href{https://doi.org/10.1007/978-1-4684-2001-2_9}
{in \emph{Complexity of Computer Computations}, 85-103 (Plenum Press, New York, 1972).}

\bibitem{RefBulger2003}
D. Bulger, W. P. Baritompa, and G. R. Wood, ``Implementing pure adaptive search with Grover's quantum algorithm,''
\href{https://doi.org/10.1023/A:1023061218864}
{\emph{Journal of Optimization Theory and Applications} \textbf{116}, 517--529 (2003)}.

\bibitem{RefBaritompa2005}
W. P. Baritompa, D. W. Bulger, and G. R. Wood, ``Grover's quantum algorithm applied to global optimization,''
\href{https://doi.org/10.1137/040605072}
{\emph{SIAM Journal on Optimization} \textbf{15}, 1170--1184 (2005)}.

\bibitem{RefBoyer1998}
M. Boyer, G. Brassard, P. H{\o}yer, and A. Tapp, ``Tight bounds on quantum searching,''
\href{https://doi.org/10.1002/(SICI)1521-3978(199806)46:4/5<493::AID-PROP493>3.0.CO;2-P}
{\emph{Fortschritte der Physik} \textbf{46}, 493--505 (1998)}.

\end{thebibliography}
\end{document}